\documentclass[11pt]{article}
\usepackage{amsmath,amssymb,color,graphics,epsfig,cite}

\usepackage{amsfonts}

\newcommand{\hoch}[1]{$\, ^{#1}$}

\newcommand{\be}{\begin{equation}}
\newcommand{\ee}{\end{equation}}
\newcommand{\bea}{\setlength\arraycolsep{2pt} \begin{eqnarray}}
\newcommand{\eea}{\end{eqnarray}}
\newcommand{\nn}{\nonumber}

\def\ft#1#2{{\textstyle{\frac{\scriptstyle #1}{\scriptstyle #2} } }}
\def\fft#1#2{{\frac{#1}{#2}}}

\def\0{{\sst{(0)}}}
\def\1{{\sst{(1)}}}
\def\2{{\sst{(2)}}}
\def\3{{\sst{(3)}}}
\def\4{{\sst{(4)}}}
\def\5{{\sst{(5)}}}
\def\6{{\sst{(6)}}}
\def\7{{\sst{(7)}}}
\def\8{{\sst{(8)}}}
\def\sst#1{{\scriptscriptstyle #1}}

\begin{document}

\begin{center}
{\Large {\bf Bronnikov-like Wormholes in Einstein-Scalar Gravity}}
\vspace{20pt}	

{Hyat Huang\hoch{1}, H. L\"u\hoch{2} and Jinbo Yang\hoch{3}}

\vspace{10pt}
\hoch{1}{\it College of Physics and Communication Electronics, \\
	Jiangxi Normal University, Nanchang 330022, China}

\hoch{2}{\it Center for Joint Quantum Studies and Department of Physics,\\
School of Science, Tianjin University, Tianjin 300350, China}

\hoch{3}{\it Institute for Theoretical Physics,\\
	 Kanazawa University, Kanazawa 920-1192, Japan}

\vspace{40pt}

\underline{ABSTRACT}
\end{center}
	
In this paper, we analyse the global structure of the Bronnikov wormhole, which is the most general spherically-symmetric and static solution in Einstein gravity coupled to a free massless phantom scalar.  We then introduce a scalar potential and construct a large class of exact solutions that can be viewed as generalizations of the Bronnikov wormhole.  We study the global structure and classify the parameters of these new wormholes.  For suitable parameters, some are regular black holes with a bouncing de Sitter spacetime inside the event horizon.

\vfill {\footnotesize hyat@mail.bnu.edu.cn\ \ \ mrhonglu@gmail.com \ \ \ \ j\_yang@hep.s.kanazawa-u.ac.jp}

\thispagestyle{empty}
\pagebreak

\section{Introduction}

The bending of spacetime in General Relativity (GR) provides tantalizing possibilities of spacetime geometries that inspire one's imagination: black holes, wormholes and time machines associated with the closed time-like curves.  Increasingly precise astronomical observations confirm the existence of black holes in our Universe.  These include the detections of gravitational waves from two colliding black holes \cite{Abbott:2016blz,TheLIGOScientific:2016src} and the recent direct photon picture of the black hole shadow \cite{Akiyama:2019cqa}.  Although there is not yet any evidence of wormholes and time machines, these research subjects have continued to attract attentions in the GR community \cite{Geng:2015kvs,Tsukamoto:2016,Xian:2019qmt,Bronnikov:2019sbx,Novikov:2019rus,Simpson:2018tsi,Lobo:2020kxn,Anabalon:2020loe,
Wang:2020emr,Liu:2020qia,Maldacena:2020sxe}. In addition to its traditional role as a tool for space and time travel, a wormhole may play an important role in quantum information through quantum entanglement \cite{Maldacena:2013xja,Balasubramanian:2014hda,Marolf:2015vma,Almheiri:2019qdq,Penington:2019kki,
Gao:2019nyj,Brown:2019hmk,Freivogel:2019whb}.

In 1988, Kip Thorne {\it et al.}~demonstrated all traversable wormhole in GR must violate the null energy condition \cite{Morris:1988cz}. It means that the traversable wormholes require exotic matter. The simplest such candidate may be the phantom scalar field. It turns out that phantom fields are one of the most promising candidates of dark energy in cosmology \cite{Nakonieczna:2015apa,Caldwell:1999ew}.  In particle physics, a phantom field model was proposed to study the neutrino mass generation \cite{Lu:2020hyw}. Furthermore, some studies argued that the instability problem of the phantom fields could be curable \cite{Piazza:2004df}.

The first traversable wormhole is the Ellis wormhole constructed in 1973, as an exact solution of Einstein gravity coupled to a free phantom scalar field \cite{Ellis:1973yv}. The Ellis wormhole is spherically-symmetric and static and it is a massless wormhole, symmetrically connecting two flat spacetimes. Later in same year, Bronnikov found that there was a more general wormhole solution in the theory \cite{Bronnikov:1973fh,Yazadjiev:2017twg}. Bronnikov wormhole is massive and asymmetric. In the massless limit, the Bronnikov solution reduces to the Ellis solution. Since then, many new wormhole solutions have been constructed in Einstein gravity or modified gravity theories \cite{Goulart:2017iko,Huang:2019arj,Mai:2017riq,Maldacena:2018gjk,Fu:2019vco,
Fu:2019oyc,Mustafa:2020gmc,Sharif:2020xxj,Lobo:2020jfl,Ibadov:2020btp,Liu:2020yhu,
Jusufi:2020yus,Mehdizadeh:2020nrw,Godani:2019kgy,Bronnikov:2019ugl}.

The purpose of this paper is to generalize the Bronnikov wormhole and construct a more general class of such solutions in Einstein-scalar gravity.  To do so, we first analyse the global structure of the Bronnikov wormhole.  We find that the speeds of light of the two Minkowski spacetimes connected by the wormhole are different.  In other words, there is no globally defined time such that the flow of time in each asymptotic region is the same.  This is very different from the Ellis wormhole, where the two asymptotic regions can be connected by the Poincare transformation without using the wormhole.  Thus although the Bronnikov wormhole solution involves two integration constants, the ratio of the speeds of light must be held fixed, leaving only one adjustable parameter.  Keeping this in mind, we introduce a scalar potential and obtain a large class of exact solutions that can be viewed as generalizations of the Bronnikov wormhole.  These solutions connect not only the Minkowski spacetimes, but also de Sitter (dS) and/or anti-de Sitter (AdS) spacetimes.

The paper is organized as follows. In section 2, we give a brief review of the Bronnikov wormhole and then study its global structure. In Section 3, we present the Lagrangian of the the Einstein-scalar theory and the corresponding local solutions of the Bronnikov-type wormholes. We examine and classify the parameters of the solutions in Section 4. We conclude the paper in section 5.

\section{Brief review of the Bronnikov wormhole}

\subsection{The theory and the most general local solution}

Einstein gravity minimally coupled to a massless scalar admits exact spherically-symmetric and static solutions.  When the scalar is phantomlike, the solution with appropriate integration constant describes a wormhole.  To be specific,
the Lagrangian is given by
\be
{\cal L}=\sqrt{-g}(R+\ft 1 2 (\partial \phi)^2)\,,
\ee
where the scalar field has negative kinetic energy and hence phantomlike. The most general spherically static wormhole solution in the theory is the Bronnikov wormhole. Following the convention in Ref.~\cite{Yazadjiev:2017twg}, the solution is
\bea\label{bronnikov}
&&ds^2=-h(r) dt^2+ h(r)^{-1}dr^2+R^2(r)d\Omega_{2}^2\,,\nn\\
&&h=e^{-\fft{M}{q}\phi},\qquad R^2=\fft{r^2+q^2-M^2}{h}\,,\nn\\
&&\phi=\fft{2q}{\sqrt{q^2-M^2}}\arctan(\fft{r}{\sqrt{q^2-M^2}})\,,\label{ebsol}
\eea
where $(M,q)$ are two integration constants and we should not mix the Ricci scalar $R$ and the coordinate function $R(r)$.  When $M=0$, the solution reduces to the well-known Ellis wormhole. The $q=0$ limit is more subtle and we find
\be
ds^2_{q=0} = -ds_{\rm sch}^2\,,
\ee
where $ds_{\rm sch}^2$ is the Schwarzschild metric of mass $-M$, rewritten in $r\rightarrow r-M$ coordinate, namely
\be
ds_{\rm sch}^2 = -\fft{(r+M)}{r-M} dt^2 + \fft{r-M}{r+M} dr^2 + (r-M)^2 d\Omega^2_2\,.
\ee
Note that the reality and regularity conditions imply that there is no smooth $q\rightarrow 0$ limit from the general wormhole solution which requires that $q> M$.

\subsection{Some global properties}

The Bronnikov wormhole (\ref{ebsol}) connects two asymptotic Minkowski spacetimes that are not symmetric.  As $r\rightarrow \pm \infty$, we have
\be
ds^2 \rightarrow -c_\pm ^2 dt^2 + dR^2 + R^2 d\Omega_2^2\,,
\ee
where
\be\label{speed}
c_\pm = e^{\mp\frac{\pi  M}{2\sqrt{q^2-M^2}}}\,,
\ee
In other words, the speeds of light at the two asymptotic flat regions are not the same and hence cannot be simultaneously set to 1 by rescaling the time or changing the unit.  Thus the asymptotic flat worlds are intrinsically different and the wormhole is a tunnel connecting them.  This should be contrasted with the Ellis wormhole where $c_+/c_-=1$, in which case, the wormhole can be a shortcut connecting two locations of the same asymptotic Minkowski world.  For two given Lorentz invariant worlds, the Bronnikov wormhole connecting them contains only one free parameter, with
\be
\fft{c_-}{c_+} = e^{\frac{\pi  M}{\sqrt{q^2-M^2}}}\qquad \hbox{fixed.}\label{fixed}
\ee
In other words, the dimensionless ratio $q/M>1$ is fixed.  On the other hand, the $c_+ c_-=1$ is coincidental; a scaling of time coordinate can turn the product to be any constant value.

\begin{figure}[ht]
	\centering
	\includegraphics[width=6cm]{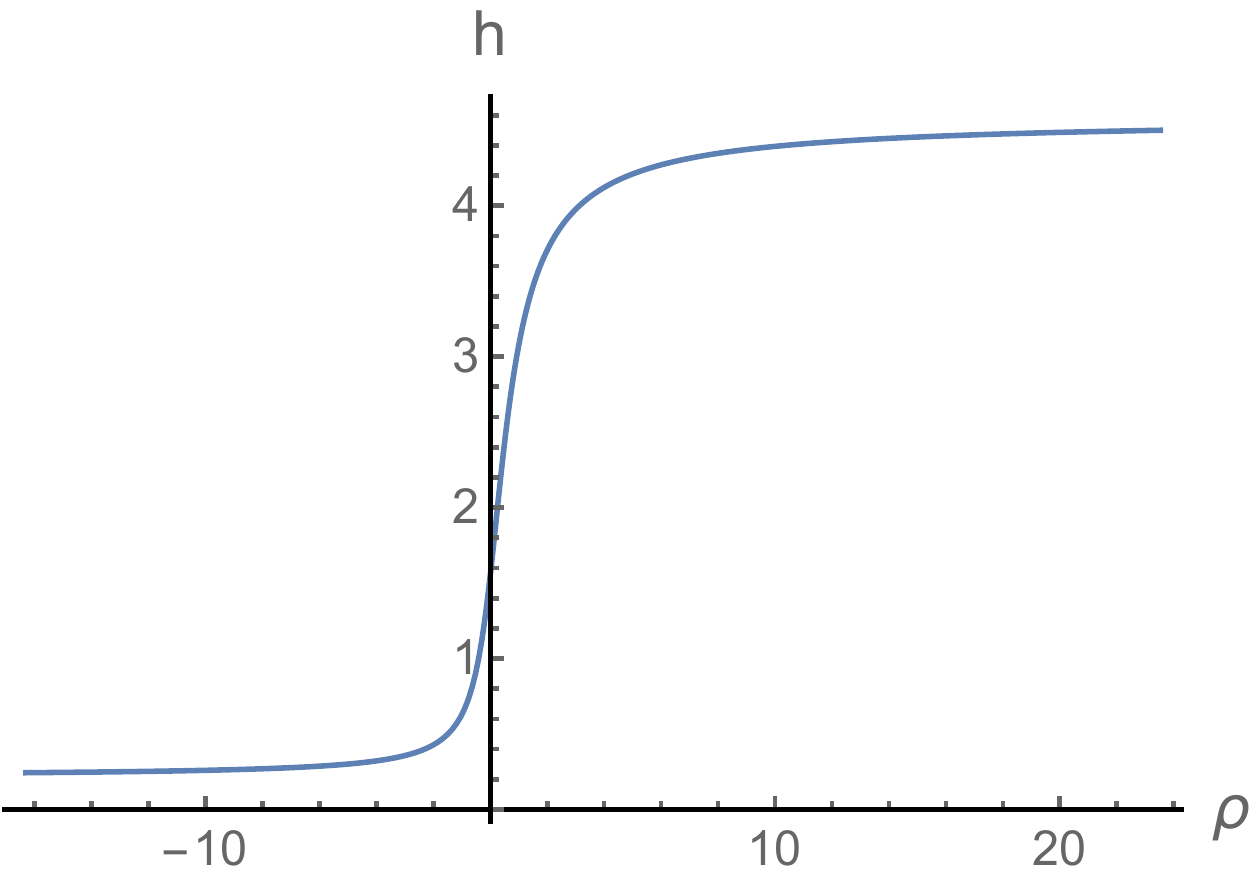}
	\caption{\small \it Here is the Bronnikov wormhole of $M=-3,q=5$, corresponding to $M_\pm=\pm 3e^{\pm3\pi/8}$ and $c_\pm=e^{\pm 3\pi/8}$. This is an asymptotically flat wormhole with the two sides of wormhole throat $(\rho=0)$ asymmetric.}
	\label{fig:bro}
\end{figure}

To understand the wormhole solution better, it is advantageous to write the metric in the more standard Ellis wormhole type of coordinates, namely
\be\label{wormmetric}
ds^2 = -  h(\rho) dt^2 + \fft{d\rho^2}{\tilde f(\rho)} + (\rho^2 + a^2) d\Omega_{2}^2\,,
\ee
where $a$ is the radius of the wormhole throat, given by
\be
a^2=R^2_{\rm min}=q^2 \exp\left(-\frac{2 M}{\sqrt{q^2-M^2}} \tan ^{-1}\left(\frac{M}{\sqrt{q^2-M^2}}\right)\right)\,,
\ee
located at $r=-M$ of the original radial coordinate. The Ellis-wormhole radius $\rho$ runs from $-\infty$ to $\infty$, connecting the two asymptotic Minkowski spacetimes.  For our purpose, it is sufficient to obtain the metric functions $h(\rho)$ and $\tilde f(\rho)$ at large $|\rho|$.  We find for $\rho\rightarrow \pm \infty$, the leading falloffs are
\bea
h &=&  c^2_\pm\Big(1 + \frac{2 M e^{\pm \frac{\pi  M}{2 \sqrt{q^2-M^2}}}}{\rho}\Big) + {\cal O}(\rho^{-3})\,,\nn\\
\tilde f &=& 1 + \frac{2 M e^{\pm\frac{\pi  M}{2 \sqrt{q^2-M^2}}}}{\rho} + {\cal O}(\rho^{-2})\,.
\eea
Note that the wormhole throat radius $a$ first appears in $h$ in the $1/\rho^3$ falloff in the asymptotic expansion, whilst it is in $1/\rho^2$ term in $\tilde f$.  The mass of the wormhole measured in the two asymptotic regions are not the same, given by
\be\label{massb}
M_\pm = \mp M e^{\mp \frac{\pi  M}{2 \sqrt{q^2-M^2}}}\,,\qquad \fft{M_+}{M_-} = -\fft{c_+}{c_-}\,.
\ee
In Fig.~\ref{fig:bro}, we plot $h$ as a function $\rho$, for parameters $(M,q)=(-3,5)$, giving rise to $M_\pm=3e^{\pm3\pi/8}$ and $c_\pm=e^{\pm 3\pi/8}$.

We see from Fig.~\ref{fig:bro} that for the parameter $M<0$, the function $h$ is a monotonically increasing function of $\rho$, approaching $c_\pm^2$ as $\rho\rightarrow \pm \infty$.  The gravitational force is thus always pointing to the negative $\rho$ direction. In other words, the wormhole is attractive in the $\rho> 0$ world, whilst it is repulsive in the $\rho<0$ world. This is consistent with the fact that $M_+>0$, but $M_-<0$. These interesting feartures could also be found in the charged Ellis wormhole \cite{Huang:2019arj}. It is also important to note that for static observers at the two asymptotic $\rho\rightarrow \pm \infty$ regions, the experienced time flow is different, namely
\be
(\Delta t)^+ = \fft{c_-}{c_+} (\Delta t)^-\,.
\ee
Specifically, the time flows faster in the world where the wormhole is attractive, than the other side where the wormhole is repulsive.  Thus the wormhole can be used as a time machine, provided that such two asymptotic Minkowski universes exist.

As we have seen, the solution appears to have two independent parameters $(M,q)$; however, the mass and wormhole radius of the Bronnikov wormhole that connects them must be related so that $c_+/c_-$ is fixed.  We can thus introduce a fixed dimensionless {\it positive} parameter
\be
\gamma=\fft{\sqrt{q^2-M^2}}{q}\le 1\,.\label{brongamma}
\ee
The wormhole solution \eqref{bronnikov} becomes
\bea\label{ds}
&&\phi=\ft{2}{\gamma}\arctan(\ft{r}{\gamma q}),\qquad R^2=e^{-\sqrt{1-\gamma^2}\phi}(r^2+\gamma^2q^2),\qquad h=e^{\sqrt{1-\gamma^2}\phi}.
\eea
In the next section, we construct a more general class of wormholes with the fixed $\gamma$ parameter, in some Einstein-scalar theories.

\section{Generalization by a scalar potential}

\subsection{Construction}

In this section,  we consider a more general phantom theory by introducing a scalar potential $V(\phi)$. The Lagrangian is given by
\be\label{lag}
{\cal L}=\sqrt{-g}(R+\ft 1 2 (\partial \phi)^2-V)\,.
\ee
The covariant equations of motion associated with the variations of $g^{\mu\nu}$ and $\phi$ are respectively given by
\bea\label{eom}
\Box\phi =-\ft{\partial V}{\partial \phi} \,,\qquad E_{\mu\nu} \equiv R_{\mu\nu}-\ft 12 R g_{\mu\nu}-T_{\mu\nu}^{\phi}=0\,,
\eea
where
\be
T_{\mu\nu}^{\phi} =-\bigg(\ft 12\partial_\mu\phi \partial_\nu \phi-\ft 14 g_{\mu\nu} (\partial \phi)^2\bigg)-\ft 1 2g_{\mu\nu}V.
\ee
To construct a spherically-symmetric and static solution, we take the general ansatz,
\be
ds^2=-h dt^2+h^{-1}dr^2+R^2(r)d\Omega_2^2,\qquad \phi=\phi(r).\label{generalsol}
\ee
Substituting this into the equations \eqref{eom} gives three independent ordinary differential equations:
\bea\label{eomm}
E^t_t=0&:&-4+4hR'^2+R^2(2V-h\phi'^2)+4R(h'R'+2hR'')=0,\nn\\
E^r_r=0&:&-4+4Rh'R'+4hR'^2+R^2(2V+h\phi'^2)=0,\nn\\
E_{\rm sphere}=0&:&R(-h\phi'^2+2(V+h''))+4(h'R'+hR'')=0.
\eea
It is straightforward to verify that the scalar equation of motion is automatically satisfied provided that the above are all satisfied.

We now use the reverse-engineering technique to construct new wormholes and determine the scalar potential. We assume that the scalar $\phi$ and the metric function $R$ of the new wormholes take exactly the same forms as the Bronnikov wormhole, given by (\ref{ds}), namely
\be\label{broansatz}
\phi=\ft{2}{\gamma}\arctan(\ft{r}{\gamma q})\,,\qquad R^2=e^{-\sqrt{1-\gamma^2}\phi}(r^2+\gamma^2q^2)\,,
\ee
where we take $q\geq 0$ in this paper without lose generality. Under the assumption \eqref{broansatz}, the $E^t_t=0$ and $E^r_r=0$ equations become equivalent. We can therefore choose one of them  and  $E_{\rm sphere}=0$  in \eqref{eomm} to obtain
\bea
h''&=&-\ft{2}{(r^2+\gamma^2q^2)^2}\bigg(\phi(r^2+\gamma^2 q^2)+(2qr\sqrt{1-\gamma^2}-r^2+q^2(\gamma^2-2))h)\bigg)\,,\\
V&=&\ft{2}{(r^2+\gamma^2 q^2)^2}\bigg((q^2(\gamma^2-2)-r^2+2q r\sqrt{1-\gamma^2})h+(r^2+\gamma^2q^2)(\phi+(1\sqrt{1-\gamma^2}-r)h')\bigg).\nn
\eea
It turns out that these two equations can be solved exactly and we obtain $h(r)$ and $V(r)$ involving two integration constants $(\alpha,\beta)$.  We can then use the $\phi(r)$ expression to write $V$ in terms of the scalar. We obtain
\bea
\label{vv}
V&=&e^{\sqrt{1-\gamma^2}\phi}\beta\Big(-3\gamma\sqrt{1-\gamma^2}\sin(\gamma\phi)+(3\gamma^2-2)\cos(\gamma\phi)-2\Big)\nn\\
&&+e^{-\sqrt{1-\gamma^2}\phi}\alpha \Big(3\gamma\sqrt{1-\gamma^2}\sin(\gamma\phi)+(3\gamma^2-2)\cos(\gamma\phi)-2\Big).
\eea
We therefore obtain the full Einstein-scalar theory (\ref{lag}). The corresponding analytical solution is then given by \eqref{broansatz} together with
\bea
h&=&\ft{1}{2}e^{-\sqrt{1-\gamma^2}\phi}\bigg(2e^{2\sqrt{1-\gamma^2}\phi}+\alpha \gamma^2(r^2+q^2\gamma^2)\nn\\
&&\qquad+\beta e^{2\sqrt{1-\gamma^2}\phi}\gamma^2\big(r^2-q^2(7\gamma^2-8)+4 q r \sqrt{1-\gamma^2}\big)\bigg).
\label{hsol}
\eea
Thus we see that we have a total of four parameters $(\alpha, \beta,\gamma,q)$ in the system. The parameters $(\alpha,\beta,\gamma)$ appear in the Lagrangian and hence they cease to be integration constants of the solutions.  Only $q$ is the integration constant. As we have emphasized earlier, this does not necessarily reduce the generality in the Bronnikov wormholes. In what follows, we shall analyse the Einstein-scalar theory and the global properties of the solutions.

\subsection{The scalar potential}

We now analyse the scalar potential \eqref{vv}.  We compute
\bea\label{dV}
\ft{\partial V}{\partial\phi}&=&2e^{-\sqrt{1-\gamma^2}\phi}\cos(\ft{1}{2}\gamma\phi)\times\nn\\
&&\bigg(2(\alpha-e^{2\sqrt{1-\gamma^2}\phi}\beta)\sqrt{1-\gamma^2}\cos(\ft{1}{2}\gamma\phi)-(\alpha+e^{2\sqrt{1-\gamma^2}\phi}\beta)\gamma\sin(\ft{1}{2}\gamma\phi)\bigg).
\eea
Obviously, the above vanishes when
\be
\phi=(\ft{1}{2}+n)\ft{2\pi}{\gamma}\qquad n=0,\pm1,\pm2,\pm3,....
\ee
These give the stationary points of $V$. We take $n=0$ and $n=-1$ and obtain two special stationary points
\be
\phi_+=\ft{\pi}{\gamma},\qquad \phi_-=-\ft{\pi}{\gamma}.
\ee
As we shall discuss in the next section, these two stationary points play an important role. By convention, we define
\be
V(\phi_\pm) = -3\gamma^2 \bigg(\alpha  e^{\mp\fft{\pi\sqrt{1-\gamma^2}}{\gamma}} + \beta  e^{\pm\fft{\pi\sqrt{1-\gamma^2}}{\gamma}}\bigg)\equiv 2\Lambda_\pm\,.
\ee
We shall see that $\Lambda_\pm$ are nothing but the effective cosmological constants in two asymptotic regions.

\begin{figure}[ht]
	\centering
	\includegraphics[width=6cm]{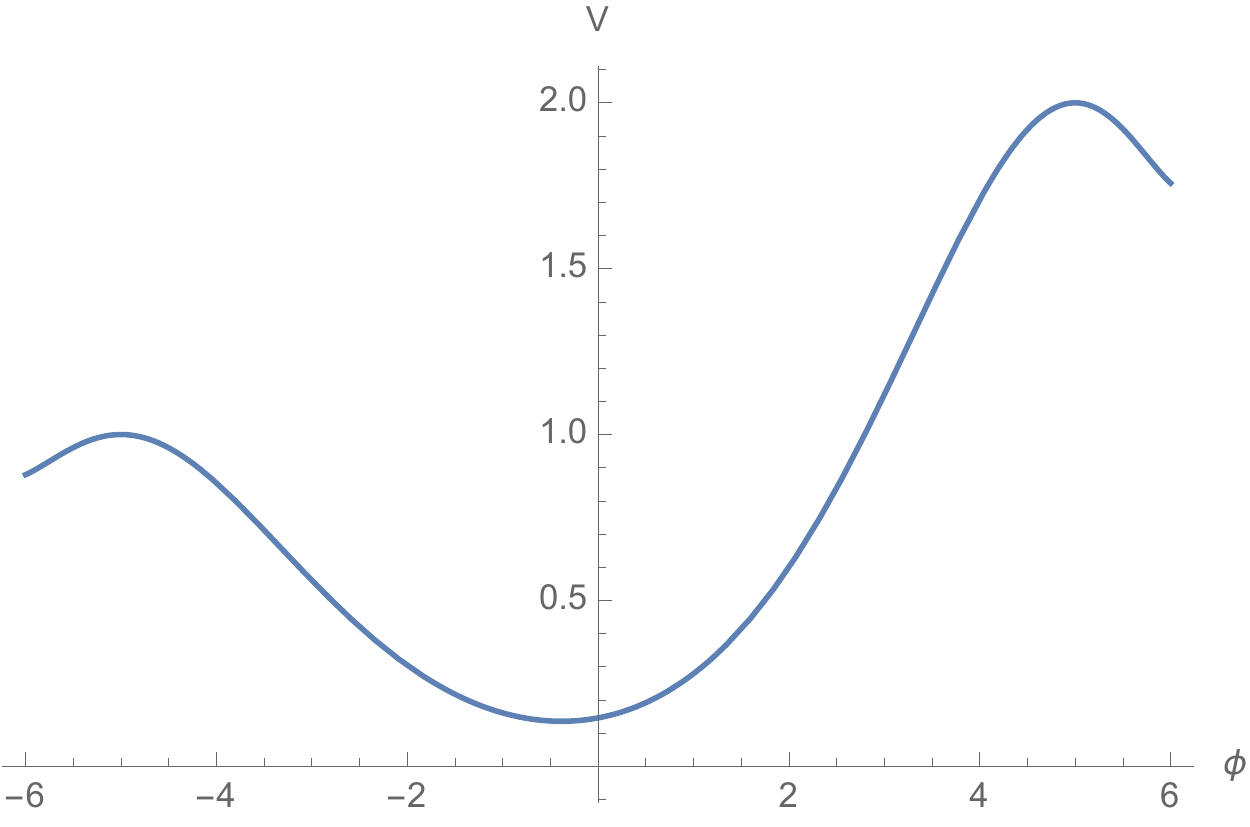}\ \ \
	\includegraphics[width=6cm]{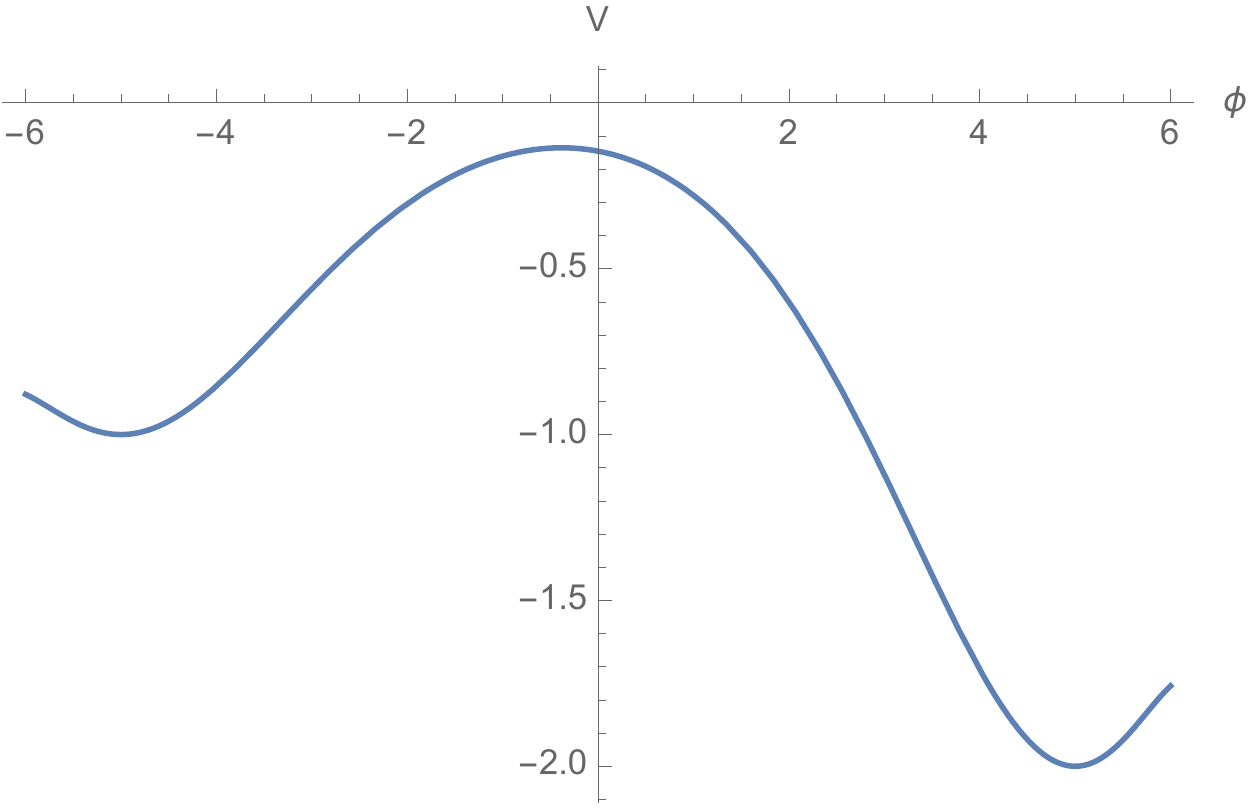}
	\caption{\small \it The left scalar potential has $\Lambda_+=1, \Lambda_-=\ft{1}{2}, \gamma=\ft{\pi}{5}$, with a minimum sandwiched between two maxima; The right one has $\Lambda_+=-1, \Lambda_-=-\ft{1}{2}, \gamma=\ft{\pi}{5}$, with a maximum sandwiched between two minima.}
	\label{fig:vphi}
\end{figure}

It is straightforward to compute
\be\label{cos}
\ft{\partial^2V}{\partial\phi^2}|_{\phi=\phi_+}=-\ft{2}{3}\Lambda_+, \qquad \ft{\partial^2V}{\partial\phi^2}|_{\phi=\phi_-}=-\ft{2}{3}\Lambda_-.
\ee
In the case of $\Lambda_+$ and $\Lambda_-$ are both positive, $\phi_+$ and $\phi_-$ give two local maximums, and there must exist at least one stationary point between the minima at $(\phi_-,\phi_+)$. Likewise, if $\Lambda_+$ and $\Lambda_-$ are both negative, $\phi_+$ and $\phi_-$ are two local minimums. Then there will be at least one stationary point between the two maxima.  These two situations are illustrated in Fig.~\ref{fig:vphi}.

\begin{figure}[ht]
	\centering
	\includegraphics[width=6cm]{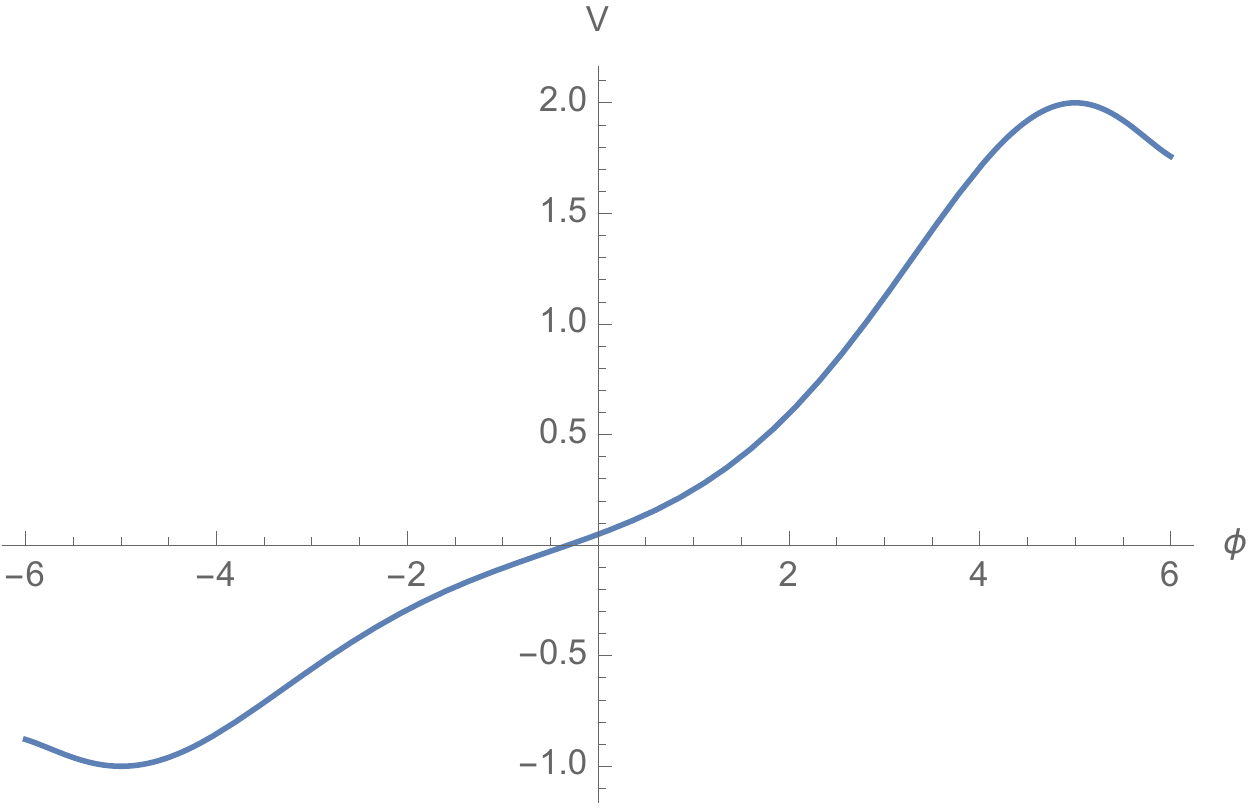}
	\caption{\it \small There is no further stationary point within $(\phi_-,\phi_+)$ when $\Lambda_+\lambda_-<0$. In this plot, we set $\Lambda_+=1, \Lambda_-=-\ft{1}{2}, \gamma=\ft{\pi}{5}$.}
	\label{vphi3}
\end{figure}

However, if $\Lambda_+$ and $\Lambda_-$ have different signs, the two stationary points then give one maximum and one minimum. In this case, there may not be further stationary point within $(\phi_-,\phi_+)$, as illustrated in Fig.~\ref{vphi3}.

\section{Wormholes and regular black holes}

\subsection{general properties}

For the Einstein-scalar theory (\ref{lag}) with the scalar potential \eqref{vv}, we have constructed the exact spherically-symmetric and static solution of the form \eqref{generalsol} where the scalar and metric functions are given by \eqref{broansatz} and \eqref{hsol}.  The radial coordinate $r$ runs from minus infinity to plus infinity, for which the scalar runs from one stationary point $\phi_-$ to $\phi_+$.  In order to study the two asymptotic structure, it is instructive to use $R(r)$ as the radial coordinate and write the metric as
\be
ds^2=-h(R) dt^2+f(R)^{-1}dR^2+R(r)^2d\Omega_2^2\,,
\ee
where $f(R)=h(R)R'(r)^2$.  Since $R(r)$ has a non-vanishing minimum, in the $R$ coordinate, $f(R)$ will have a single zero while $h(R)$ has no zero.  Thus the $R$ coordinate will not describe the full spacetime region; nevertheless, it is useful to describe the asymptotic structure and read off the mass.  In the $R\rightarrow \pm \infty$ regions, the coordinate $r$ and $R$ are related by
\be
r\sim e^{\pm\ft{\sqrt{1-\gamma^2}\pi}{2\gamma}}R -\ft{2q^3\sqrt{1-\gamma^2}}{3R^2}- \ft{q^2}{2R}e^{\mp\ft{\sqrt{1-\gamma^2}\pi}{2\gamma}}-q\sqrt{1-\gamma^2} +
e^{\mp\ft{\sqrt{1-\gamma^2}\pi}{2\gamma}}+\cdots\,.
\ee
Thus in the $R\rightarrow \pm \infty$ asymptotic region, the solutions become the (A)dS vacua
\be
ds^2 = -c_\pm ^2 \Big(-\ft13{\Lambda_\pm}R^2+1\Big) dt^2 + \fft{dR^2}{\Big(-\ft13{\Lambda_\pm}R^2+1\Big)} + R^2 d\Omega_2^2\,,\qquad \phi=\phi_\pm\,,
\ee
where
\be
c_\pm = e^{\mp\fft{2\sqrt{1-\gamma^2}\pi}{\gamma}}\,.
\ee
Note that the speeds of light in the two vacua take the same forms as those in the Bronnikov wormhole discussed earlier. The masses of the solutions in both asymptotic regions can be obtained by examining the asymptotic falloffs and we find
\be
M_\pm=\pm\ft13 {q\sqrt{1-\gamma^2}\big(3-2q^2\gamma^2\beta(3\gamma^2-4)\big)}
e^{\mp\fft{\sqrt{1-\gamma^2}\pi}{2\gamma}}.
\ee

As in the case of the Bronnikov wormhole, the best coordinate to describe the full wormhole geometry is the Ellis-wormhole coordinate, namely \eqref{wormmetric}, where the the radial coordinate $\rho$ runs smoothly from $-\infty$ to $+\infty$, correcting the two asymptotic spacetimes.  The radius $a$ of the wormhole throat $(\rho=0)$ is given by
\be
a^2=R^2_{\rm min}=q e^{-\fft{\sqrt{1-\gamma^2}}{\gamma}\arctan(\fft{\sqrt{1-\gamma^2}}{\gamma})},
\ee
located at $r=q\sqrt{1-\gamma^2}$ of the origin coordinate. In the next, we shall study the global structures in detail depending on the various choices of the parameters.

\subsection{Symmetric (A)dS wormhole}

We first consider $\gamma=1$, in which case, the scalar potential \eqref{vv} is
\be
V=\ft{2}{3}\Lambda^*(\cos\phi-2)\,,
\ee
where $\Lambda^*=-\fft32(\alpha+\beta)$ is a constant. The corresponding solution reduces to
\bea
ds^2&=& -h dt^2+h^{-1}dr^2+(r^2+q^2) d\Omega^2_2,\nn\\
\phi&=&2\arctan(\ft{r}{q}),\quad
h=1-\ft{1}{3}\Lambda^*(r^2 + q^2)\,.
\eea
Thus we see that the solution is already naturally in the Ellis-wormhole ansatz with the wormhole throat at $r=0$ and wormhole radius $q$.  As $r$ runs from $-\infty$ to $+\infty$, the scalar runs from $\phi_-=-\pi$ to $\phi_+=\pi$, with $V_\pm=\Lambda^*$.  The solution describes a symmetric AdS wormhole, and was a spectial case of those obtained in Ref.~\cite {Bronnikov:2005gm,Zhang:2014sta,Huang:2019lsl}. This can be viewed as a direct (A)dS generalization of the Ellis wormhole and the mass is zero.

\subsection{Asymmetric (A)dS wormhole }

\begin{figure}[ht]
	\centering
	\includegraphics[width=6cm]{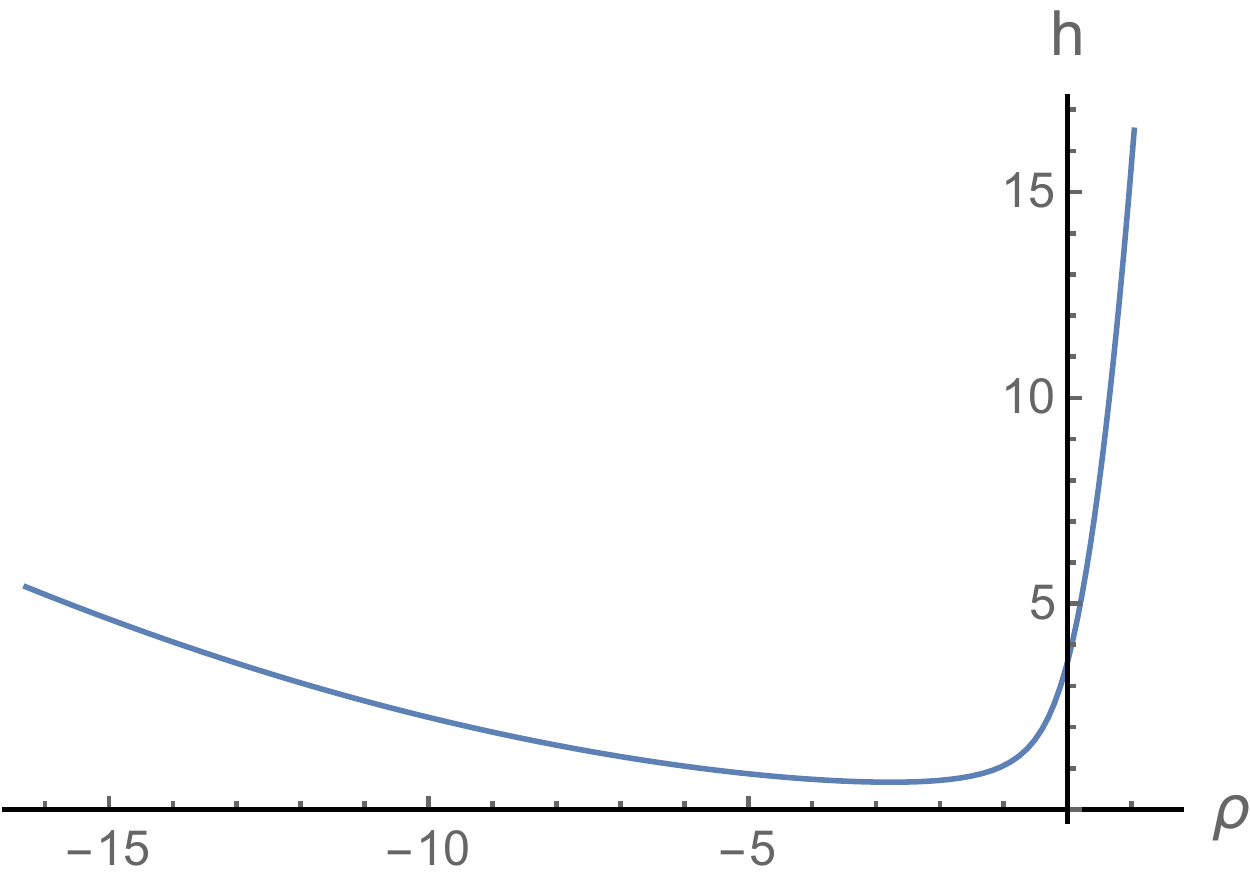}
	\caption{\small \it The AdS wormhole: the parameters are set in  $\alpha=0, \beta=1,q=1,\gamma=0.9$. This is a asymptotic AdS wormhole that the two sides of wormhole throat are asymmetric. The effective cosmological constant of the two asymptotic regions are $\Lambda_-=-0.295$ and $\Lambda_+=-6.182$ respectively.}
	\label{fig:wh2h}
\end{figure}

We now consider the more general $0<\gamma< 1$ case. We first consider that the two asymptotic spacetimes are both
AdS or dS. In other words, the effective cosmological constants of the two sides satisfy
\be
\Lambda_+ \Lambda_->0\,.
\ee
The easiest way to achieve this is to set $\alpha=0$, which leads to
\be\label{hwormhole}
h=\ft{1}{2}e^{\sqrt{1-\gamma^2}\phi}\bigg(\beta \gamma^2 r^2+(4 \beta q\gamma^2\sqrt{1-\gamma^2}) r+2+\beta q^2\gamma^2(8-7\gamma^2)\bigg).
\ee
For $\beta>0$, the effective cosmological constants
\be
\Lambda_\pm = -\ft32\beta\gamma^2e^{\pm\fft{\pi\sqrt{1-\gamma^2}}{\gamma}}
\ee
are both negative, giving rise  to the two asymptotic  AdS spacetime. It can be verify that there is no real root in $h=0$ for $0<\gamma<1$ and $\beta>0$. The the solution describes a wormhole connecting two AdS worlds with different cosmological constants. The function $h$ in terms of the Ellis-wormhole radius $\rho$ is depicted in Fig.~\ref{fig:wh2h}

\begin{figure}[ht]
	\centering
	\includegraphics[width=6cm]{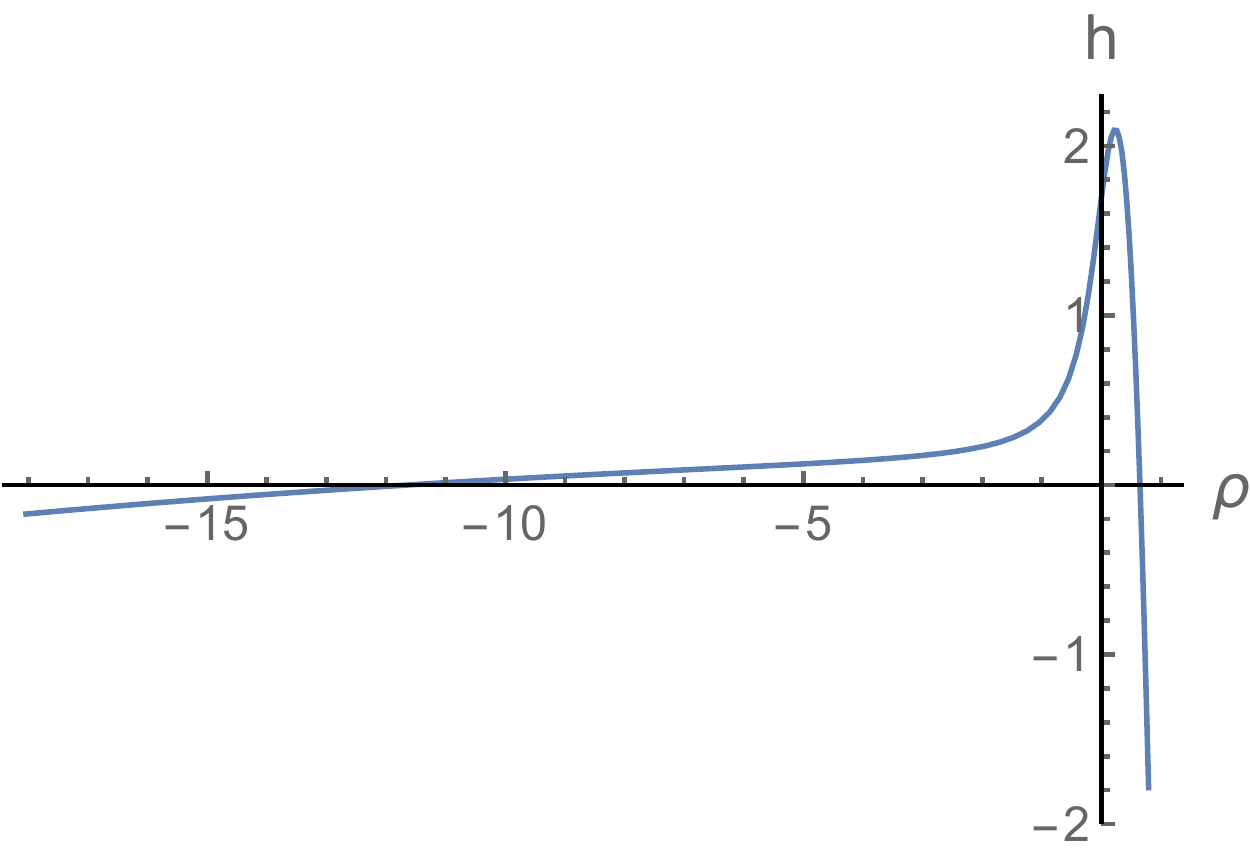}
	\caption{\small \it The asymptotic dS spacetime: the parameters are set in  $\alpha=0, \beta=-\ft{225}{752},q=\ft{5}{6},\gamma=\ft{8}{10}$. The two cosmological horizons locate at $r_-=-4$ and $r_+=2$ respectively.  The effective cosmological constant in asymptotic region of $\rho<0$ is $\Lambda_-=0.034$ whlie in the $\rho>0$ side is $\Lambda_+=3.788$.}
	\label{fig:ds}
\end{figure}

On the other hand, the parameter choice $\beta<0$ gives rise to two asymptotic dS spacetimes. There is a cosmic horizon associated with the the dS spacetime in each side, given by
\be
r_\pm=-2 \sqrt{1-\gamma^2} q \pm \sqrt{-\fft{2}{\beta\gamma^2} - q^2 (4-3\gamma^2)},\nn\\
\ee
The reality condition requires that
\be
q\in [0,\ft{\sqrt{2}}{\sqrt{-\beta\gamma^2(4-3\gamma^2)}}).
\ee
The wormhole throat is located at $r=q\sqrt{1-\gamma^2}$ $(R=R_{\rm min})$. The asymmetric dS wormhole is depicted in Fig.~\ref{fig:ds}.

Note that if we turn on the $\alpha$ parameter while maintaining $\Lambda_+\Lambda_->0$, The wormhole solutions are analogous and we shall not give further discussions.

\subsection{Connecting (A)dS to flat spacetime}

\begin{figure}[ht]
	\centering
	\includegraphics[width=6cm]{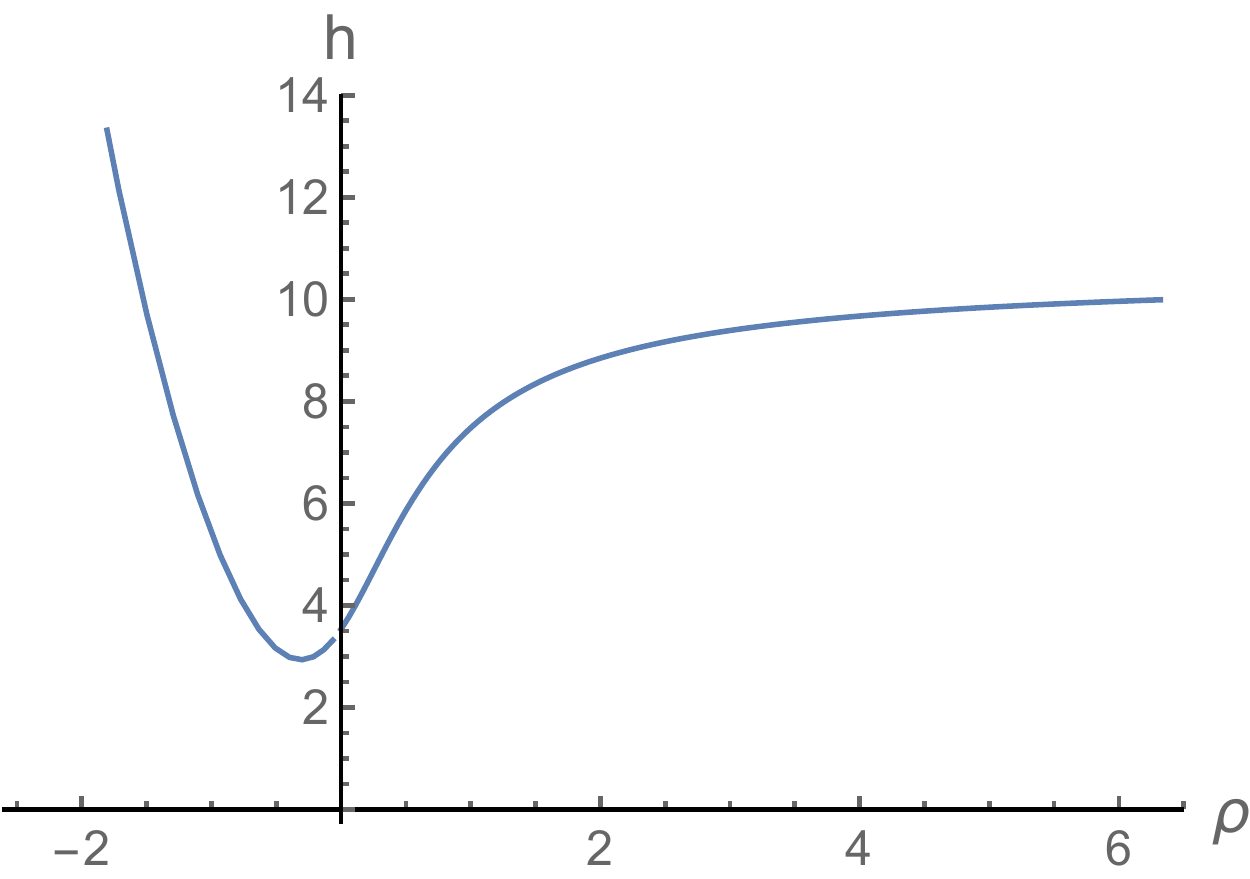}
	\caption{\small \it This is the function $h(\rho)$ of the wormhole connecting AdS to flat spacetimes, with $\alpha=-e^{\fft{2\pi \sqrt{1-\gamma^2}}{\gamma}}\beta=-11.1318, \beta=-0.1,q=1,\gamma=0.8$. The effective cosmological constant of the two asymptotic regions are $\Lambda_-=-140.927$ and $\Lambda_+=0$ respectively.}
	\label{fig:wh2hflat}
\end{figure}

We now consider the case with one the effective cosmological constant vanishes.  Without loss of generality, we choose $\Lambda_+=0$.  This can be achieved by setting $\alpha=-e^{\fft{2\pi \sqrt{1-\gamma^2}}{\gamma}}\beta$, in which case, we have
\be
\Lambda_-=\ft{3}{2}\beta\gamma^2 e^{-\ft{\pi\sqrt{1-\gamma^2}}{\gamma}}(e^{\ft{4\pi\sqrt{1-\gamma^2}}{\gamma}}-1)\,.
\ee
For $\beta<0$, the solution is wormhole connecting the AdS spacetime at $\rho\rightarrow -\infty$ to the flat spacetime at $\rho\rightarrow +\infty$.
The function $h$ in terms of the Ellis wormhole coordinate $\rho$ is depicted in Fig.~\ref{fig:wh2hflat}.

\begin{figure}[ht]
	\centering
	\includegraphics[width=6cm]{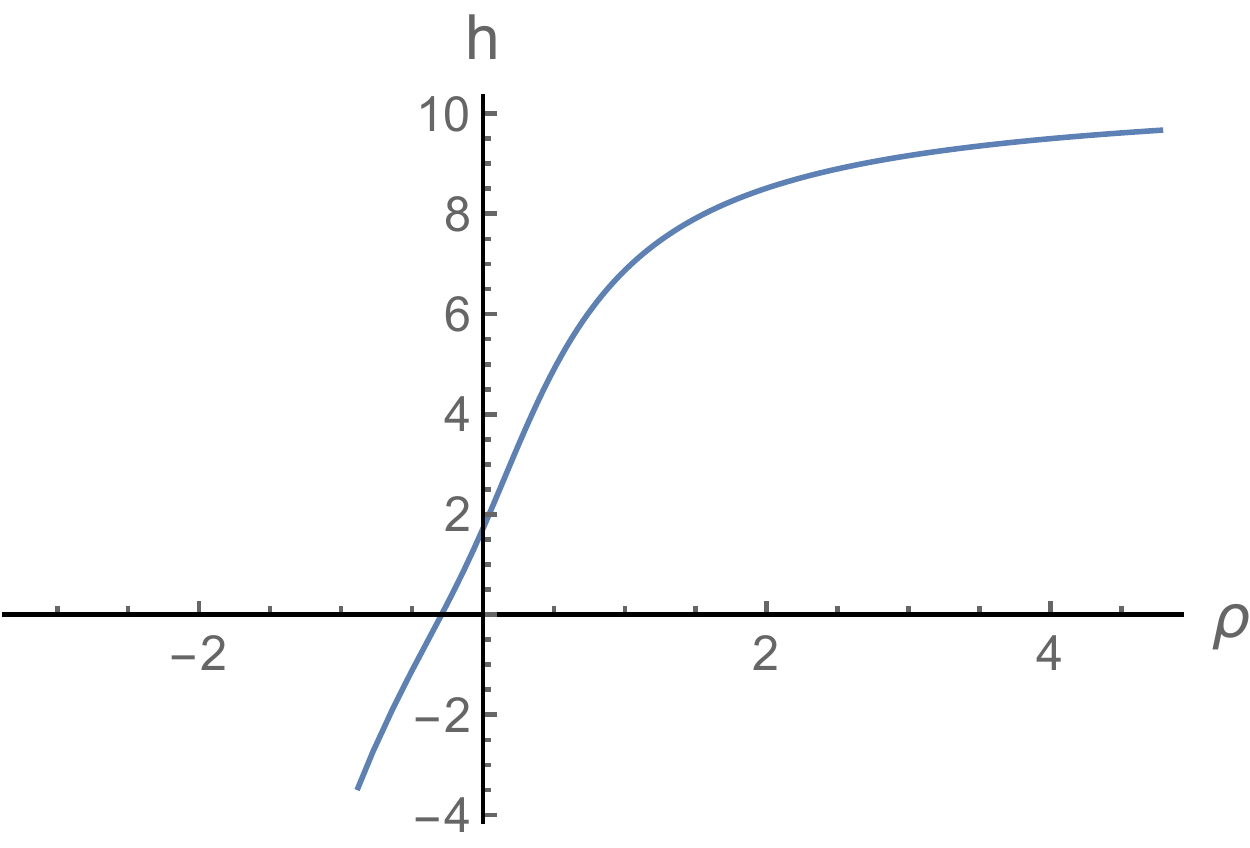}
	\includegraphics[width=6cm]{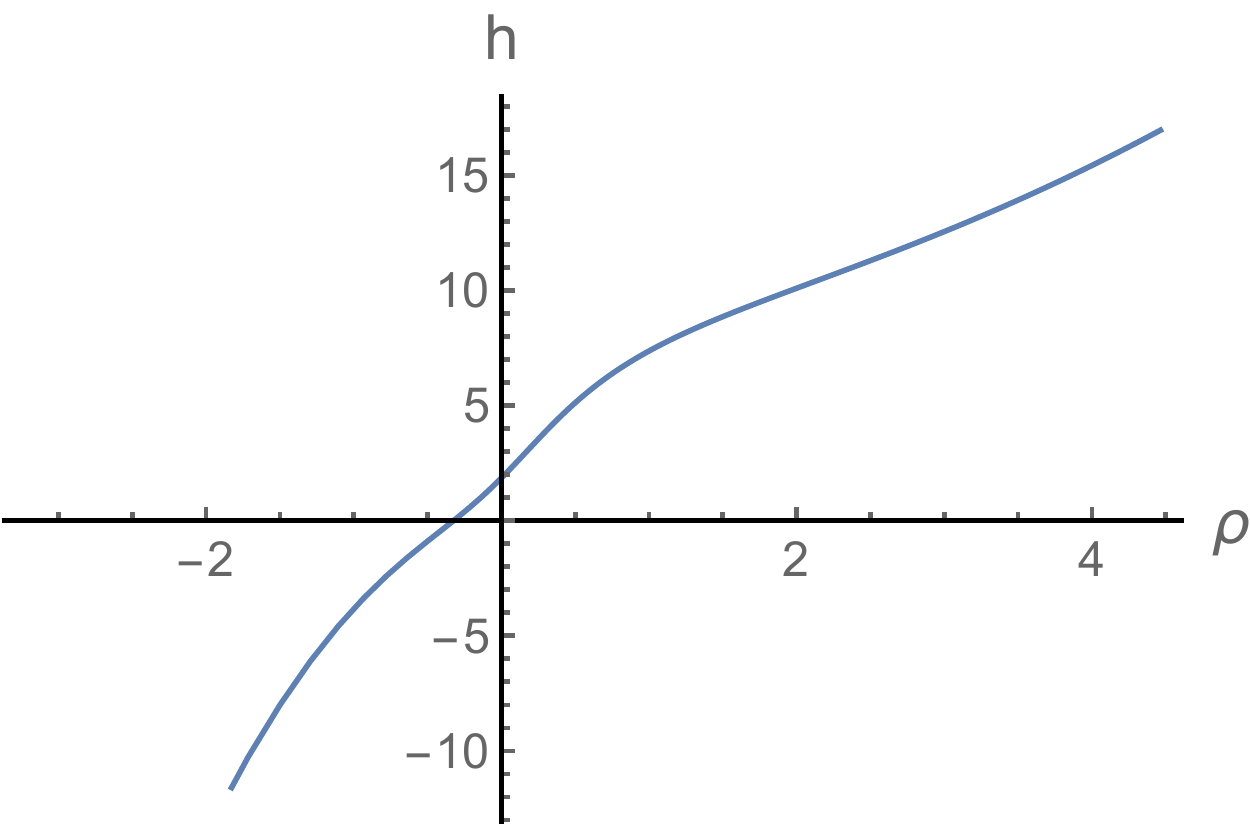}
	\caption{\small \it The left one is $h$ of the wormhole connecting the dS and flat spacetimes, with parameters $\beta=0.1, \alpha=-11.1318, \gamma=0.8, q=1$ then $\Lambda_-=140.927$. The cosmic horizon is located at $\rho_H^c=-0.291$. The right one of the wormhole connecting dS to AdS spacetimes, with $\beta=0.1, \alpha=-10, \gamma=0.8, q=1$ and the cosmic horizon is located at $\rho_H^c=-0.322$.  The effective cosmological constants of the two asymptotic regions are $\Lambda_-=126.597$ and $\Lambda_+=-0.129$ respectively.}
	\label{fig:bh2}
\end{figure}

When $\beta>0$, the effective cosmological constant $\Lambda_-$ is positive and hence there is an cosmic horizon in the $\rho<0$ region, as can be seen in the left graph of Fig.~\ref{fig:bh2}.  In the right graph, we also changed the $\alpha$ parameter so that the wormhole connects the dS spacetime ($\Lambda_-<0$) to the AdS spacetime ($\Lambda_+>0$).

\subsection{A bouncing universe inside a black hole}

\begin{figure}[ht]
	\centering
	\includegraphics[width=6cm]{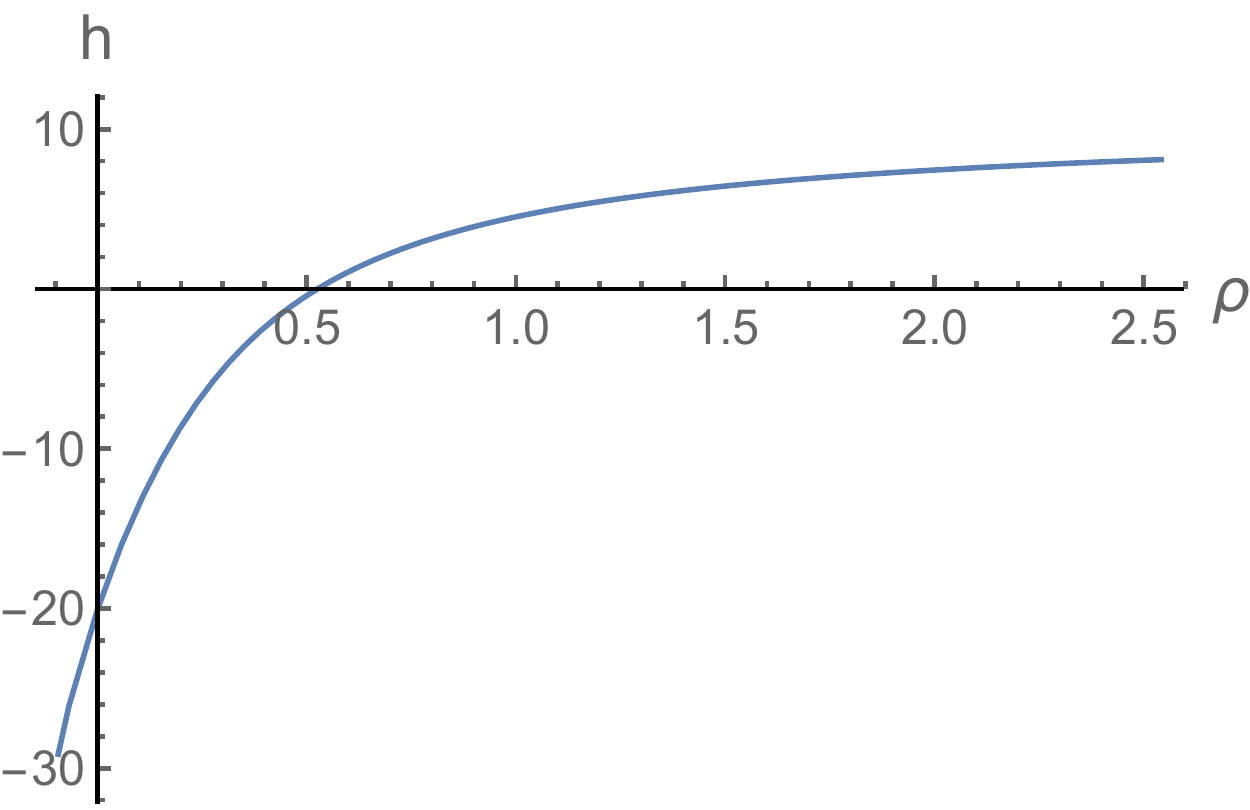}
	\includegraphics[width=6cm]{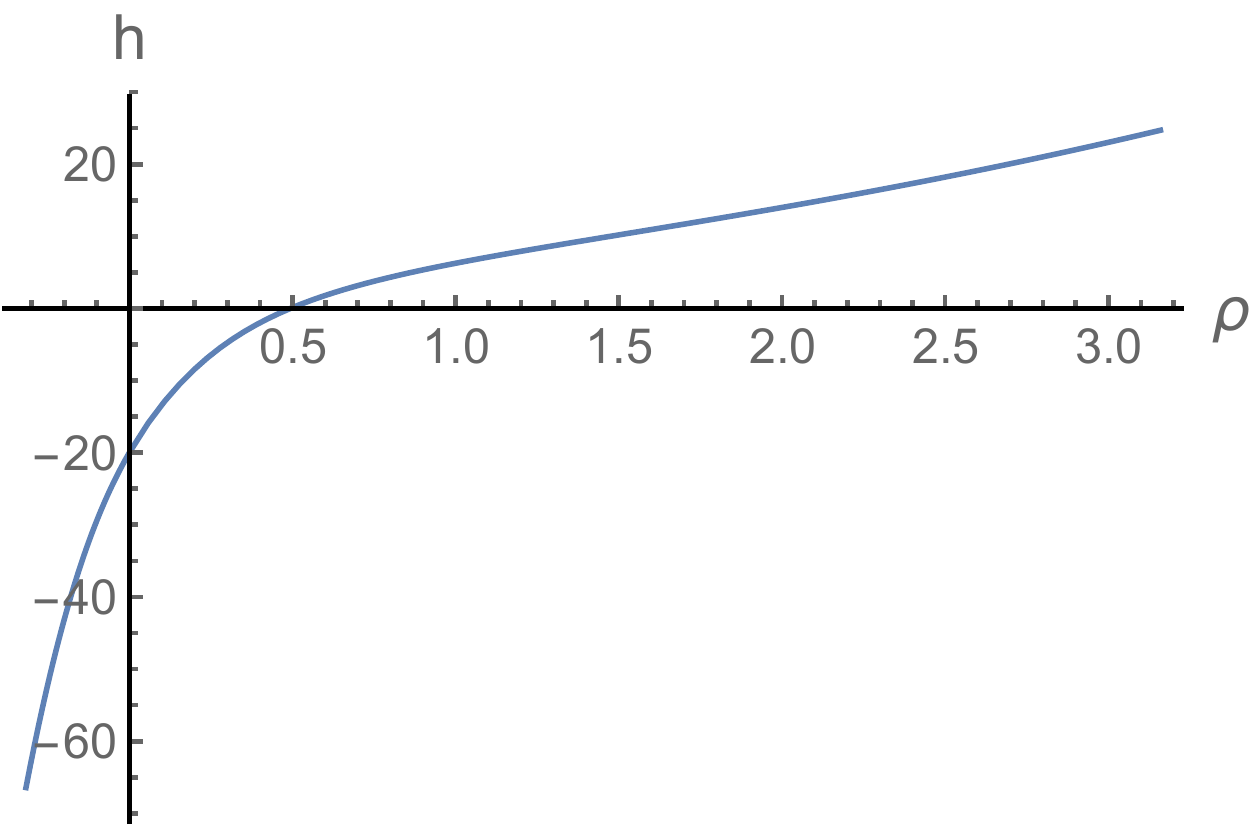}
	\caption{\small \it For $\alpha=-1113.18, \beta=10, q=0.5, \gamma=0.8$, then $\Lambda_-=14092.7$ we have an asymptotically flat black hole with horizon located at $\rho=0.526$;  For $\alpha=-1112.878,\beta=10,q=0.5, \gamma=0.8$, we have an asymptotically AdS black holes with horizon $\rho_H=0.496$. The effective cosmological constants of the two asymptotic regions are $\Lambda_-=14029.4$ and $\Lambda_+=-0.569$ respectively. In both cases, inside the horizon is a cosmological bouncing dS universes, with no curvature singularity.}\label{fig:bh3}
\end{figure}

In the previous subsection, for positive $\Lambda_-$, the root $\rho_H^c<0$ of the vanishing $h(\rho)$ is a cosmic horizon in the $\rho<0$ world of the $\rho=0$ wormhole throat.  If we can adjust the parameters such that the root becomes positive, then it becomes a black hole horizon $\rho_H$ in the $\rho>0$ world.  We find that such parameters do exist and the result is plotted in Fig.~\ref{fig:bh3}.

``Inside'' the black hole, $\rho$ becomes timelike and the solution is cosmological and there is no singularity at $\rho=0$.  Instead, $\rho=0$ now describes a cosmological bounce that into a new universe with different speed of light.

\section{Conclusions}

In this paper, we first studied the global structure of the Bronnikov wormhole that is the spherically-symmetric and static solution of Einstein gravity coupled to a free massless phantom scalar. We demonstrated that the wormhole connected two Minkowski worlds with different speeds $c_\pm$ of light.  Although in each world, one can scale the time coordinate such that the speed of light is unit; however, when they are connected by the wormhole, there is no globally defined time such that the speeds of light can be set to unit simultaneously.  The two worlds are thus not connected by the Poincare transformations, but only through the wormhole. This should be contrasted with the Ellis wormhole, which can be a shortcut for travelling between two locations in the same Minkowski spacetime.

The Bronnikov wormhole involves two parameters $(M,q)$, parameterizing the radius of the wormhole throat and the mass.  The wormhole is asymmetric and the mass $M_\pm$ in the two sides of the wormhole are different, with one positive and one negative.  In other words, the wormhole appears attractive in one world, but repulsive in the other.  By contrast, the Ellis wormhole is symmetric and massless. An intriguing feature of the Bronnikov wormhole is that it can be used as a time machine since the time flow of the two asymptotic worlds are different, with the time flowing faster in the world where the wormhole is attractive.

In the Bronnikov wormhole universe, for the two fixed asymptotic Minkowski spacetimes, the wormhole involves actually only one integration constant since $c_+/c_-$ must be fixed.  In other words, the dimensionless parameter $\gamma$ \eqref{brongamma} must be fixed.  This led to the second part of the this paper, where we constructed a general class of the scalar potential involving the parameter $\gamma$.  This allowed us to construct a more general class of the Bronnikov type of exact wormhole solutions that connect not only the flat spacetimes, but also the (A)dS spacetimes.  We analysed the global structures of these wormhole solutions, and classified the parameters for different types of asymptotic structures.  We also obtained examples of regular black holes with a bouncing dS universe inside its horizon.

The general wormholes involve four parameters, but the effective cosmological constants $\Lambda_\pm$ of both worlds and the parameter that determines ratio of the speeds of light are all fixed by the Einstein-scalar theory itself.  Thus our exact solutions involve only one integration constant, with no independent scalar hair parameter. It is of interest to study how the new scalar hair would affect the wormhole properties.  Furthermore, it is interesting to examine how our discussion can be generalized to more general Einstein-Maxwell-scalar theories.

\section*{Acknowledgement}

H.H.~is supported by the Initial Research Foundation of Jiangxi Normal University. 
H.L.~is supported in part by NSFC (National Natural Science Foundation of China) Grant No.~11875200 and No.~11935009.
J.Y.~is supported by the China Scholarship Council and the Japanese Government (Monbukagakusho-MEXT) scholarship.

\end{document}